\documentclass[manuscript]{aastex631}

\newcommand\aastex{AAS\TeX}

\usepackage{amsmath}
\usepackage{bm}

\def\vec#1{\ensuremath{\bm{#1}}}

\defcitealias{2023arXiv230103100P}{P23} 
 
\received{}
\revised{}
\accepted{}

\shorttitle{\aastex\ sample article}
\shortauthors{Guo et al.}

\begin{document}

\title{Influence of the Lower Atmosphere on Wave Heating and Evaporation in Solar Coronal Loops}

\correspondingauthor{Mingzhe Guo, Timothy Duckenfield}
\email{mingzhe.guo@kuleuven.be, tim.duckenfield@northumbria.ac.uk}

\author{Mingzhe Guo}
\affiliation{Centre for mathematical Plasma Astrophysics, Department of Mathematics, KU Leuven, Celestijnenlaan 200B, B-3001 Leuven, Belgium}

\author{Timothy Duckenfield}
\affiliation{Department of Mathematics, Physics and Electrical Engineering, Northumbria University, Newcastle upon Tyne, NE1 8ST, UK}

\author{Tom Van Doorsselaere}
\affiliation{Centre for mathematical Plasma Astrophysics, Department of Mathematics, KU Leuven, Celestijnenlaan 200B, B-3001 Leuven, Belgium}

\author{Konstantinos Karampelas}
\affiliation{Centre for mathematical Plasma Astrophysics, Department of Mathematics, KU Leuven, Celestijnenlaan 200B, B-3001 Leuven, Belgium}

\author{Gabriel Pelouze}
\affiliation{Université Paris-Saclay, CNRS, Institut d’astrophysique spatiale, 91405, Orsay, France}

\author{Yuhang Gao}
\affiliation{School of Earth and Space Sciences, Peking University, Beijing, 100871, China}
\affiliation{Centre for mathematical Plasma Astrophysics, Department of Mathematics, KU Leuven, Celestijnenlaan 200B, B-3001 Leuven, Belgium}

\begin{abstract}
We model a coronal loop as a three-dimensional 
magnetic cylinder in a realistic solar atmosphere 
that extends from the chromosphere to the corona.
Kink oscillations, believed ubiquitous in the solar corona,
are launched in the loop.
Heating is expected due to the dissipation of wave energy 
at small structures that develop from 
the Kelvin-Helmholtz instability 
induced by kink oscillations.
Increases in temperature and internal energy can be observed in the coronal counterpart of the driven loop. 
With the presence of thermal conduction,
chromospheric evaporation can also be seen.
Although the volume averaged temperature and density changes 
seem slight ($\sim4\%$ relative to a non-driven loop),
the enthalpy flow from the lower atmosphere redistributes
the density and temperature in the vertical direction,
thus enhancing the dissipation of wave energy in the corona.
The efficient heating in the coronal counterpart of the loop 
can complement the thermal conductive losses shown 
in the current model and thus maintain the internal energy
in the corona.

\end{abstract}

\section{INTRODUCTION} 
\label{sec_intro}

More and more observations have confirmed the 
omnipresence of kink waves
in coronal loops
\citep[e.g.,][for recent reviews]{2020ARA&A..58..441N, 2021SSRv..217...73N}
and thus bring kink waves into the forefront of coronal heating consideration \citep[see][for recent reviews]{2012RSPTA.370.3193D,2015RSPTA.37340261A,2020SSRv..216..140V}. 
In the heating process, 
effective dissipation of wave energy 
can be achieved when the Lundquist number
and Reynolds number,
 which are typically very large
($\sim10^{12}$) in the solar corona,
become very small \citep[][]{2015ApJ...803...43S}.
Thus 
efficient heating relies on
the occurrence 
of small spatial scales.
Generally,
kink waves are believed to suffer resonant absorption 
\citep[see][for a review]{2011SSRv..158..289G} that
transfers wave energy from collective kink modes to
local Alfv\'en waves with the occurrence of 
transverse inhomogeneity around the loop boundary 
\citep[e.g., ][]{2020ApJ...904..116G}. 
The converted localized azimuthal Alfv\'en waves undergo 
phase mixing \citep[e.g.,][]{1983A&A...117..220H}. 
The subsequent phase mixed Alfv\'en waves can enhance 
the Kelvin-Helmholtz instability
\citep[KHI,][]{1984A&A...131..283B}, 
which is induced by velocity shear between loops and the 
background corona,
thereby facilitating the wave dissipation by further generating 
smaller scales.
This scenario is first demonstrated by \citet[][]{1994GeoRL..21.2259O} in coronal loop models.
Recently, 
numerical progress confirmed that transverse waves in coronal
loops can induce the KHI, 
forming the TWIKH rolls
\citep[Transverse Wave Induced Kelvin-Helmholtz rolls, see e.g.,][]{2008ApJ...687L.115T,2015ApJ...809...72A,2017A&A...604A.130K, 2017A&A...602A..74H,2019ApJ...870...55G, 2019ApJ...883...20G,2021ApJ...908..233S}.
Non-linear damping of kink waves associated with the KHI
has been investigated
numerically by e.g., \citet{2016A&A...595A..81M} 
and analytically by e.g., 
\citet{2021ApJ...910...58V}.
During such a process, 
the energy of collective kink waves can dissipate at such 
turbulent small structures,
and quantitative evaluations of heating effects have been made by e.g., \citet[][]{2017A&A...604A.130K,2019A&A...623A..53K,2019ApJ...870...55G,2019ApJ...883...20G}. 
Recent numerical progress shows that the heating effects 
induced by kink waves can balance the radiative loss in the 
solar corona \citep{2021ApJ...908..233S,2022ApJ...941...85D},
which shed new light on understanding coronal heating
from the perspective of waves.

The influence of chromospheric evaporation on the heating
of coronal loops was first considered by 
\citet{1998ApJ...493..474O} based on resonant kink waves.
However,
the aforementioned investigations associated with 
TWIKH rolls mainly focus on the coronal 
counterpart of loops.
In reality,
the lower atmosphere can act as a mass and energy reservoir
for physical processes in the corona.
In the process associated with impulsive heating,
nanoflares \citep[][]{1988ApJ...330..474P} for instance,
enthalpy flux plays an important role in the thermal evolution of coronal loops
\citep[e.g.,][]{2010ApJ...710L..39B,2012ApJ...752..161C}.
Regarding the AC models,
Ohmic heating has been confirmed by both observations 
\citep{2007A&A...471..311V} and numerical simulations 
\citep{2019A&A...623A..53K,2019ApJ...883...20G} in coronal
loops.
In such a scenario,
wave heating is expected to happen near loop footpoints,
where the current density that is essential for resistive dissipation
has maximum values at loop ends
\citep{2007A&A...471..311V}.
In this way,
the cooler chromosphere can be heated straightforwardly.
Given this,
a natural question arises:
Can the heating near loop footpoints trigger chromosphere 
evaporation?
The scenario of evaporation is 
usually discussed when chromospheric materials are heated 
during such energy release processes as solar flares 
\citep[e.g.,][]{1985ApJ...289..425F,2006ApJ...638L.117M, 2018ApJ...856...34T}.
Although the energy dissipation of waves is not so fierce
as flares,
gentle chromosphere evaporations are still expected 
\citep{1985ApJ...289..414F}.
The influence of such evaporation flows on the efficiency of 
wave heating in the corona remains to be evaluated
in recent models.
Recent progress by \citet{2020A&A...635A.174V} investigated 
the influence of chromosphere evaporation on the phase mixing 
of Alfv\'en waves in a two-dimensional coronal loop model.
The density shell around the loop boundary that is essential for energy
dissipation of Aflv\'en waves seems not significantly changed in the modest heating process.
However, wave energy dissipation happens over almost the whole
deformed cross-section of loops when it comes to transverse oscillations 
\citep[e.g.,][]{2018A&A...610L...9K,2019ApJ...870...55G,2019ApJ...883...20G}.
It remains to be seen how chromosphere evaporation influences wave energy dissipation
in coronal loops deformed by kink oscillations.

In this paper,
we aim to investigate the influence of the lower atmosphere
on coronal heating effects.
The paper is organized as follows. 
Section~\ref{sec_models} describes the model, 
including the equilibrium and numerical setup. 
In Section~\ref{sec_results},
we present the numerical results. 
Section~\ref{sec_sum} summarizes our findings, 
ending with some further discussion.

\section{NUMERICAL MODEL}
\label{sec_models}

\begin{figure}
	\centering
	\gridline{\fig{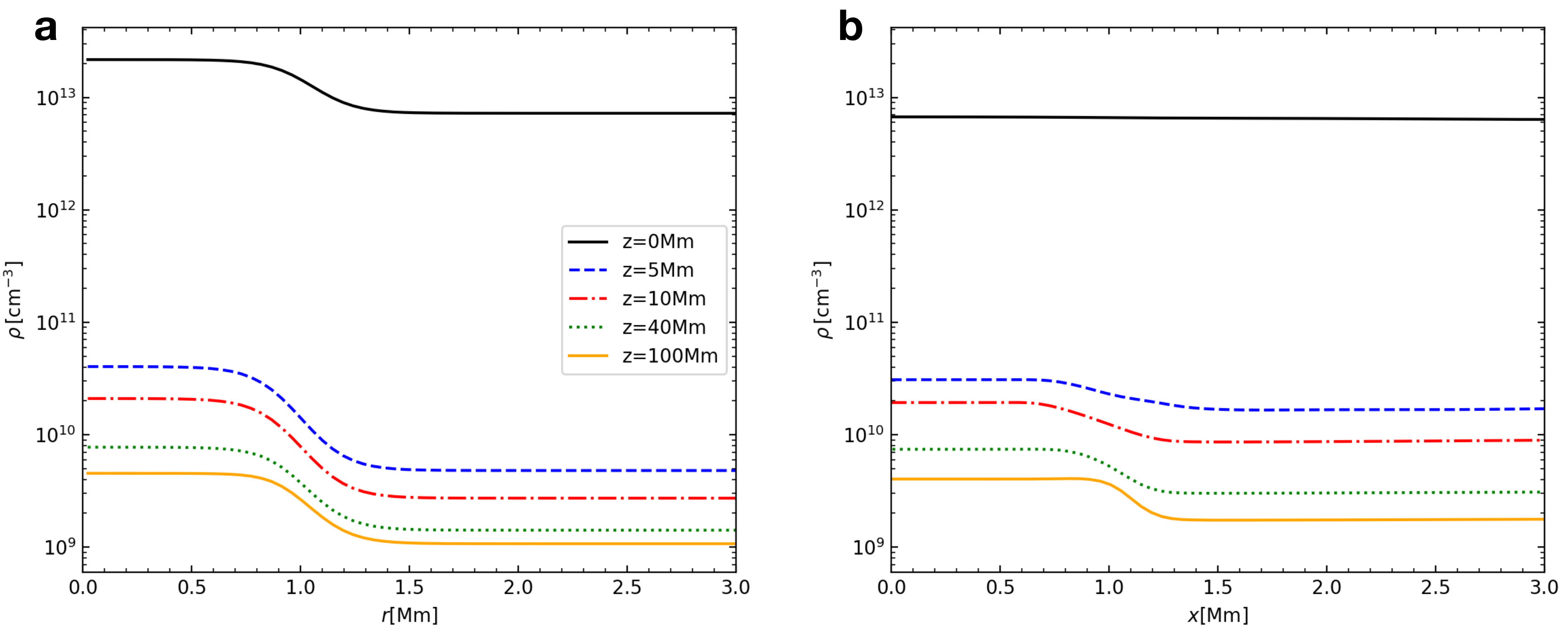}{0.9\textwidth}{}
	}
	\caption{Transverse distribution of the density along the $r$-direction in the 2D simulation (left) and $x$-direction in the 3D simulation (right). The left panel represents the initial density in the 2D simulation. The right panel shows the initial state of the 3D simulation after relaxation. Different colors represent different heights.}
	\label{fig_z}
\end{figure}

\begin{figure}
	\centering
	\gridline{\fig{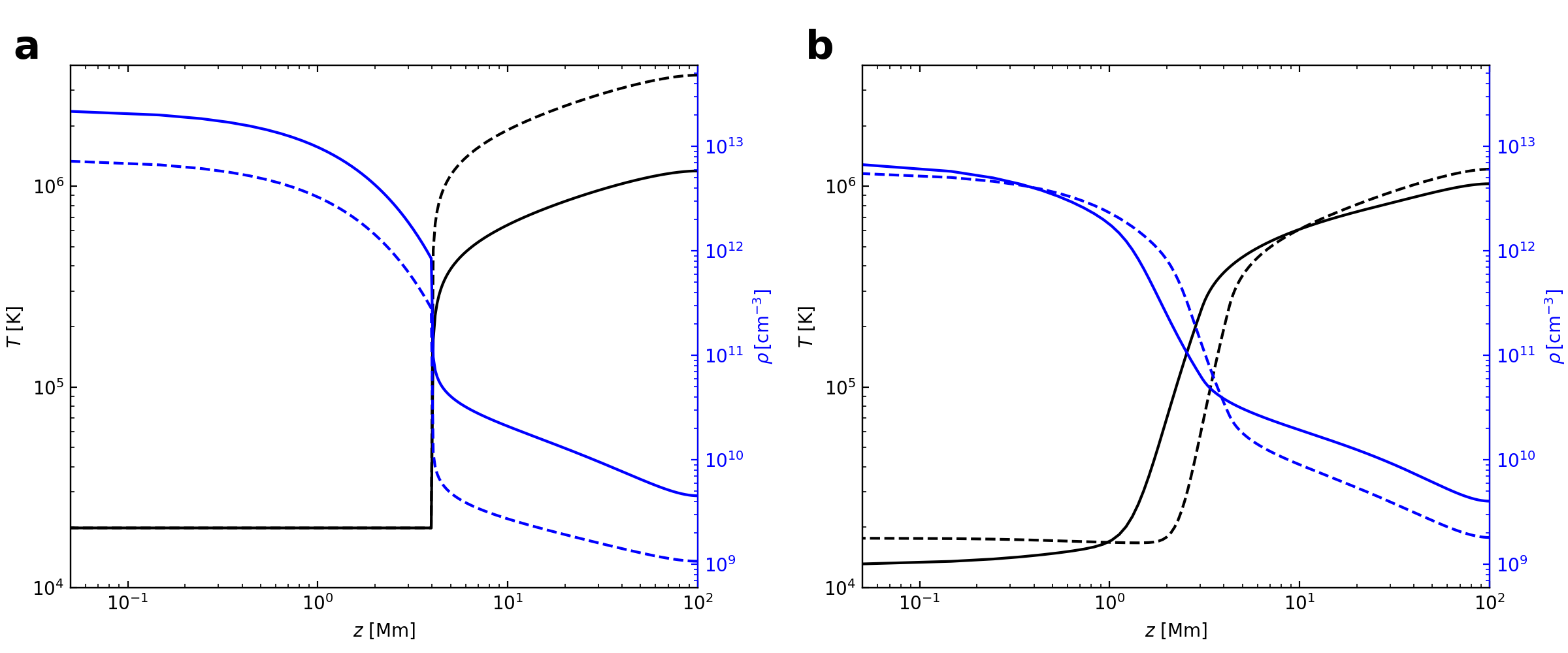}{0.9\textwidth}{}
	}
	\caption{Spatial distribution of temperature and density
along the $z$-direction. The left panel displays the initial quantities before relaxation in the 2D equilibrium. The right panel shows the initial state of the 3D simulation after relaxation. Solid and dashed lines represent the distribution at the loop axis ($x=0$, $y=0$) and the external loop region ($x=8$Mm, $y=0$), respectively. }
	\label{fig_init}
\end{figure}

We use a similar magnetic flux tube employed in \citet[][\citetalias{2023arXiv230103100P}]{2023arXiv230103100P},
with the key difference that we simulate half of a closed loop 
with both footpoints anchored in the chromosphere.
The model is initiated from a 2D hydrostatic equilibrium
in cylindrical coordinates ($r,z$).
A loop-oriented gravity 
$g(z)=-g_\odot\sin\left[\pi(L-z)/(2L)\right]$ is considered,
with $L=100$Mm being the half length of the loop.
The initial magnetic field is $\vec{B}=B_0\hat{z}$, 
with $B_0$ set to 42G.
The temperature profile, derived from \cite{2002ApJS..142..269A},
 is given by
\begin{eqnarray}
  T(r,z)
= \left\{
   \begin{array}{ll}
 T_{\rm ch},&z\le z_{\rm ch},\\
 T_{\rm ch}+\left[T_{\rm co}(r)-T_{\rm ch}\right]\left[1-\left(\displaystyle\frac{L-z}{L-z_{\rm ch}}\right)^2\right]^{0.3},&z> z_{\rm ch},
    \end{array}
  \right.
\label{eq_temp}
\end{eqnarray}
where
$T_{\rm ch}=20000$K is the temperature of the chromosphere,
and $z_{\rm ch}=4$Mm represents its thickness.
The transverse temperature profile is defined as
\begin{eqnarray}
  T_{\rm co}(r)
= T_{\rm e}+\displaystyle\frac{1}{2}\left(T_{\rm i}-T_{\rm e}\right)
   \left\{1-\tanh\left[\left(\displaystyle\frac{r}{R}-1\right)b\right]\right\},
\label{eq_temp_r}
\end{eqnarray}
where $T_{\rm i}=1.2$MK ($T_{\rm e}=3.6$MK) is the temperature inside (outside) the tube.
$R=1$Mm gives the radius of the loop,
and $b=10$ defines the thickness of the inhomogeneous layer,
corresponding to $l/R=0.6$ at the loop bottom.
The transverse distribution of the density in
Figure~\ref{fig_z}a clearly shows the variation of this layer thickness
along the height.
The initial state is not in magnetohydrostatic balance,
thus a relaxation is needed.
In this stage,
we redefine the velocity rewrite layer,
which was proposed in \citetalias{2023arXiv230103100P}
to absorb the upward flows in the loop.
With the presence of the velocity rewrite layer,
the large flow induced by the initial state,
which is not in magnetohydrostatic balance,
is weakened.
Thus the
process of reaching an equilibrium becomes 
more gradual and gentle,
ensuring the stability of the simulation.
The velocity is modified as $v'(r,z)=\alpha(t)v_i(r,z)$.
The parameter $\alpha(t)$ is defined as
\begin{eqnarray}
  \alpha (t)
= \left\{
       \begin{array}{ll}
          0.9995+0.0005\displaystyle\frac{t}{t_c}, & 0<t\le t_{\rm c},  \\
           1, & t>t_{\rm c},
        \end{array} 
  \right.
\label{eq_alpha}
\end{eqnarray}
where $t_{\rm c}=37.3{\rm ks}$ represents the critical time.
The vertical velocity in the loop is then suppressed to be
less than $1.3{\rm km s^{-1}}$
and significant quantity changes (e.g., a drop in magnetic field strength in \citetalias{2023arXiv230103100P}
\footnote{In the current model, the magnetic field undergoes 
a slight change from the initial value of 42G to around 
40G after relaxation.}) 
are avoided after the relaxation.
The relaxed state is then straightforwardly converted to 3D 
by rotating the 2D axisymmetric results.
To avoid any unphysical oscillations due to boundary changes
from the 2D cylindrical coordinate to the 3D Cartesian domain,
we allow the 3D system to relax for another 2.8ks.
The initial temperature and density distributions of
the loop axis of the 2D and 3D simulations are shown 
in Figure~\ref{fig_init}.

Note that this simulation is gravitationally stratified, 
leading to the transverse structuring of the loop changing with height. As shown in Figure~\ref{fig_z}b, 
the inhomogeneous layer thickness is smaller at the loop apex ($z=100$Mm) than at lower heights. 
One may expect this change in layer thickness to have an impact 
on the formation of KHI \citep[e.g.,][]{2016A&A...595A..81M}.

The chromosphere and transition region are included in
the current model. 
The transition region has been artificially broadened to ensure 
its resolution,
following the scheme developed by 
\citet{2001JGR...10625165L,2009ApJ...690..902L,2013ApJ...773...94M}.
As described in \citetalias{2023arXiv230103100P},
a critical temperature $T_{\rm c}=2.5\times10^5{\rm K}$ is defined,
below (above) which the parallel thermal conductivity is 
$\kappa_\parallel=\kappa_0T_{\rm c}^{5/2}$
($\kappa_\parallel=\kappa_0T^{5/2}$), 
with $\kappa_0$ being 
$5.6\times10^{-7}{\rm erg~cm^{-1}~s^{-1}~K^{-7/2}}$.
This modified thermal conductivity changes the temperature
length scale in the vertical direction,
thus leading to a significant
broadening of the transition region after relaxation,
see also \citet{2020A&A...635A.174V} and \citetalias{2023arXiv230103100P}.
Therefore, 
a coarser resolution in the vertical direction is allowed in
the current simulation. 

\begin{figure}
	\centering
	\gridline{\fig{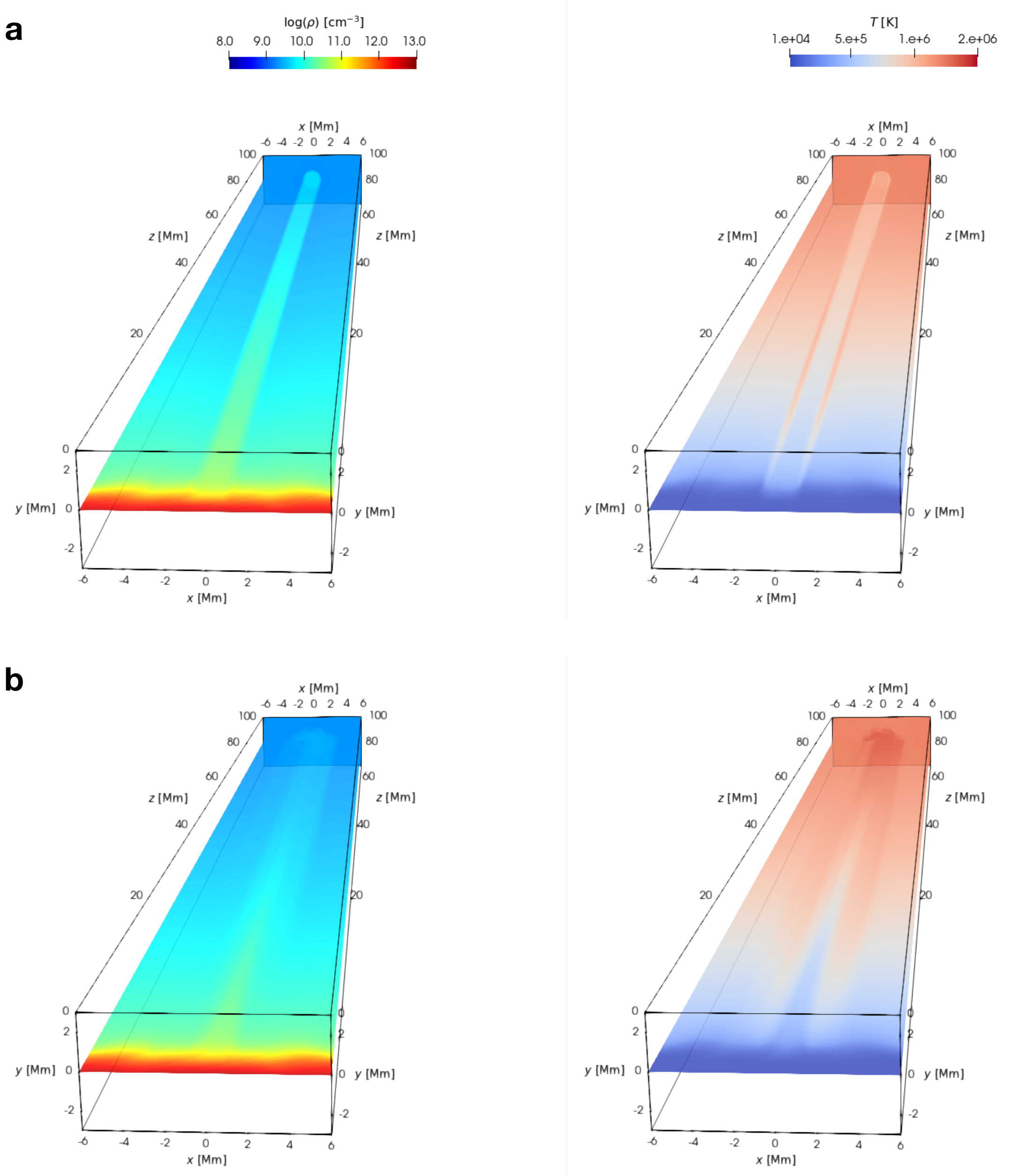}{0.9\textwidth}{}}
	\caption{Isosurfaces of the logarithm of density and temperature at the loop apex ($z=100$Mm) and $y=0$ for (a) non-driven and (b) driven model at $t=8060$s. 
	An animation of the density and temperature isosurfaces for the driven
	model is available in the online Journal. 
	The animation proceeds from $t=0\text{ to } 9340$s.}
	\label{fig_isosurface}
\end{figure}

To solve the 3D ideal MHD equations,
we employ the PLUTO code \citep{2007ApJS..170..228M}.
A piecewise linear scheme is used for spatial reconstruction.
Numerical fluxes are computed by the Roe Riemann solver,
and a second-order characteristic tracing method is used 
for time marching.
The hyperbolic divergence cleaning method is adopted to ensure 
the divergence-free nature of the magnetic field.
Anisotropic thermal conduction is included in our simulations.
In the 2D run,
the computational domain is $[0,6]{\rm Mm}\times[0,100]{\rm Mm}$.
We consider 128 uniformly spaced cells in the $r$-direction and
a uniform grid of 1024 cell points in the $z$-direction.
In the 3D case,
the simulation domain is $[-6,6]{\rm Mm}\times[-3,3]{\rm Mm}\times [0,L/2]$.
We consider uniform 1024 cell points in 
the $z$-direction
and 256 uniformly spaced cells in the $x$-direction and
128 uniform grid cells in the $y$-direction.
The resolution in the $x,y$ plane is about 46.9km.
Although increasing the transverse resolution may 
reveal more pronounced heating effects,
the main findings of the current work are not expected to be 
influenced.
On one hand, 
the changing width of the boundary layer allows the current 
resolution to sufficiently resolve fine structures away from the loop apex \footnote{Even at the loop apex, we can still observe relatively larger KHI structures, as shown in Figure~\ref{fig_isosurface}b and related animations.}.
On the other hand, 
the turbulent structures generated in the model will further extend 
the thickness of the boundary layer,
making it more feasible to reveal small scales within it.
The vertical resolution seems lower than that in
the transverse direction.
Nevertheless,
it is enough to resolve the broadened transition region that
is more than 2Mm wide after relaxation,
considering the broadening scheme 
described above.

In the 2D simulation,
an axisymmetric (outflow) boundary is employed 
at $r=0$ ($r=100$Mm).
A symmetric boundary condition is used at $z=100$Mm,
considering the symmetric property of the fundamental 
kink mode considered in this study.
At the bottom of the loop,
the density and pressure are extrapolated from the hydrostatic equilibrium.
The magnetic field is extrapolated following the zero normal 
gradient condition,
as described by \citet{2019A&A...623A..53K}.
The vertical velocity $v_z$ is set to be zero,
while the transverse velocity $v_r$ is to be outflow.
In the 3D case,
all the lateral boundaries are set to be outflow.
Boundary conditions in the $z$-direction are kept the same
as in the 2D except for the transverse velocity $v_x$ 
and $v_y$ at $z=0$, 
which are described 
by a continuous and monoperiodic driver 
\citep{2010ApJ...711..990P,2017A&A...604A.130K,2019ApJ...870...55G}.
In the current study,
the driver period is 288s,
which matches the eigenfrequency of the loop 
at the initial state.
Note that the initial eigenfrequency of the loop slightly varies
over time during the driving phase.
Due to the modest changes in the density structure 
in the current model,
the variation in eigenfrequency is not significant.
Nonetheless, a broad-band driver \citep[e.g.,][]{2019ApJ...876..100A} 
needs further study and discussion.
The amplitude of the driver is chosen to be 0 
(non-driven model)
and $4{\rm km s^{-1}}$ (driven model).
These two models are mentioned hereafter by using subscripts ``0''
and ``4'', respectively.

\section{RESULTS}
\label{sec_results}
\begin{figure}
	\centering
	\gridline{\fig{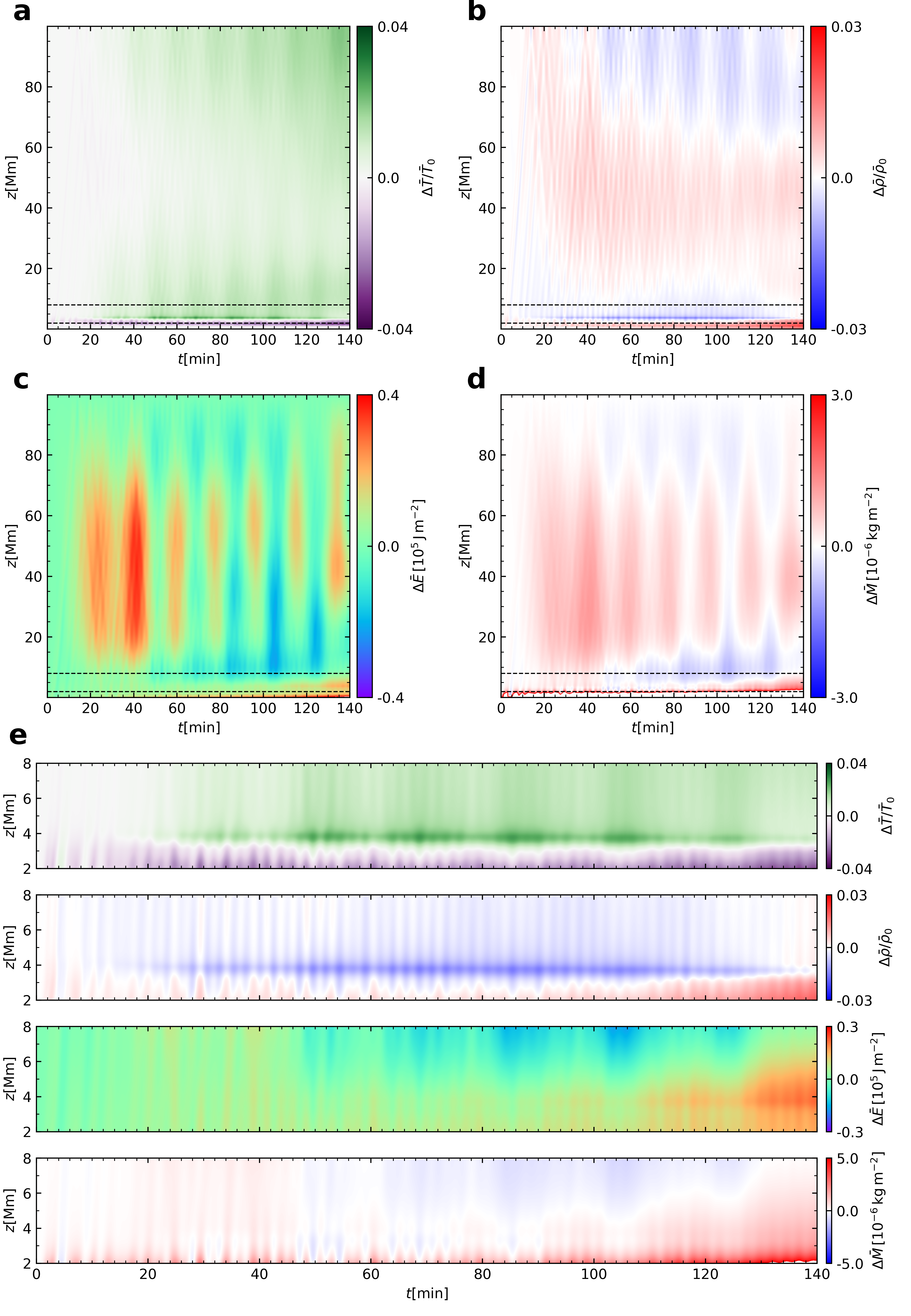}{0.84\textwidth}{}}
	\caption{Surface-averaged temperature, density, enthalpy flux, and mass flux changes relative to the non-driven case are shown versus height, with zoomed-in lower panels for the regions outlined by dashed boxes.}
	\label{fig_TrhoE}
\end{figure}
We first examine the dynamic process of the loop.
Figure~\ref{fig_isosurface} shows the isocontours of 
density and temperature
for both models at $t=8060$s when the loop has
fully evolved.
The TWIKH rolls are clearly seen from the 
density and temperature distribution at the apex of
the driven loop in Figure~\ref{fig_isosurface}b.
The core of the loop cross-section is sufficiently heated 
at the loop apex, 
even though the Ohmic dissipation is smaller
than that near the loop footpoints.
To further examine the evolution in the vertical direction,
surface averaged temperature and
density are examined along the $z$-direction,
which are given by
\begin{eqnarray}
\bar{T}(z,t)=\frac{1}{A}\int_AT{\rm d}A,
\bar{\rho}(z,t)=\frac{1}{A}\int_A\rho{\rm d}A.
\label{eq_T}
\end{eqnarray}
where $A$ represents the surface area of the region 
$-6{\rm Mm}\le x\le6{\rm Mm},-3{\rm Mm}\le y\le3{\rm Mm}$ 
at given heights. 
${\rm d}A$ is the surface element of the plane. 
Figure~\ref{fig_TrhoE}a,b show the changes in $\left(\bar{T}-\bar{T}_0\right)/\bar{T}_0$ and $\left(\bar{\rho}-\bar{\rho}_0\right)/\bar{\rho}_0$, 
which clearly reveal the variations caused by waves in 
comparison to the non-driven case.
To reveal the evaporation from the lower atmosphere,
the enthalpy flux and mass flux are also considered as
\begin{eqnarray}
\bar{E}(z,t)=\frac{1}{A}\int^t_0\int_A\left(\frac{\gamma}{\gamma-1}p\right)v_z{\rm d}A{\rm d}t,
\label{eq_en}
\end{eqnarray}
\begin{eqnarray}
\bar{M}(z,t)=\frac{1}{A}\int^t_0\int_A\rho v_z {\rm d}A{\rm d}t.
\label{eq_en}
\end{eqnarray}
Figure~\ref{fig_TrhoE}c shows the changes of 
the $z$-component of the enthalpy flux
 ($\Delta\bar{E}=\bar{E}-\bar{E}_0$).
From Figure~\ref{fig_TrhoE}a, 
we observe the expected temperature increases near the loop footpoint. This heating induces upward enthalpy flows, 
as seen in Figure~\ref{fig_TrhoE}c. 
The upward enthalpy flows are present throughout the simulation, although periodic downward flows occur after about 50min.
Figure~\ref{fig_TrhoE}b confirms the mass changes induced by the upward flow, 
as evidenced by the density increases.
A periodicity of $\sim20$min
\footnote{This period is close to the slow transit time in the coronal counterpart.} 
of the temperature profile is observed.
The density and enthalpy flux exhibit a similar
periodicity as shown in Figure~\ref{fig_TrhoE}.
This periodicity is associated with the
ponderomotive force and is attributed to slow waves,
which has also been observed in previous numerical studies
\citep[e.g.,][]{2016A&A...595A..81M, 2017A&A...604A.130K}.
This has also been confirmed by \citet{2020A&A...635A.174V}
by considering an Alfv\'en wave driver.
The excited slow oscillation can influence the vertical distribution
of the temperature profile, density distribution, and enthalpy flux.
The variations in the temperature and density profiles relative to the 
non-driven case seem slight (less than $4\%$).
This is partly because the evaporation induced by waves,
as demonstrated in \citet{2020A&A...635A.174V},
can only cause modest variations in the dramatically changing
atmosphere background in the vertical direction.
In addition,
the surface averaged values here are naturally
smaller than the local quantity values in the loop region,
which is no longer straightforward to quantify since
our current loop is fully deformed and transversely diffused than 
previous coronal models. 
Nevertheless,
the surface average procedure can still reveal the average 
variations of quantities of 
loop cross-section at a given height.
The evolutions of quantities in the domain of 
$2{\rm Mm}\le z\le8{\rm Mm}$
are zoomed in and presented in the lower panels of 
Figure~\ref{fig_TrhoE}.
We see upward enthalpy flows in this domain, 
indicating the occurrence of chromospheric evaporation.
The density drops and temperature increases around $z=4$Mm 
are the consequence of this evaporation.
Additionally,
the lower region presents a 
decrease (increase) in temperature (density) variations 
around $z=2$Mm.
Seen from Figure~\ref{fig_init},
the boundary between the chromosphere and the transition region 
shifts downward to about $z=1$Mm,
while the lower boundary of the transition region 
outside the loop is around $z=2$Mm.
This means that denser and cooler masses from the
external loop region are involved when computing the surface
averaged quantities around $z=2$Mm.

\begin{figure}
	\centering
	\gridline{\fig{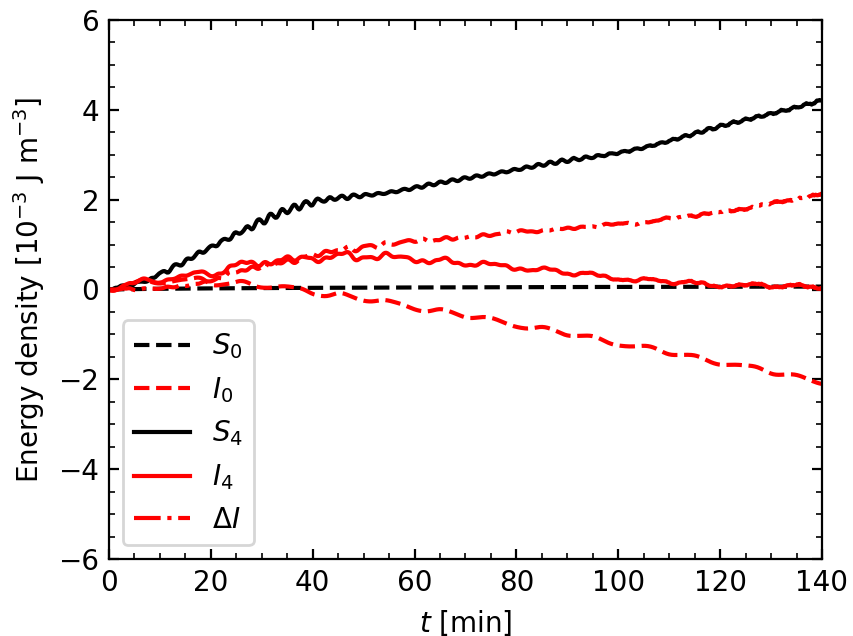}{0.7\textwidth}{}}
	\caption{Volume averaged input energy (black) and internal energy density (red) changes relative to the initial state in the coronal counterpart ($-6{\rm Mm}\le x\le6{\rm Mm},-3{\rm Mm}\le y\le3{\rm Mm},9{\rm Mm}\le z\le100{\rm Mm}$). Subscript ``0'' (``4'') denotes the non-driven (driven) model. Dot-dashed line shows the difference between the two models in the concerned domain.}
	\label{fig_energy}
\end{figure}

Heating effects induced by kink waves can be quantified by 
energy analysis 
\citep[e.g.,][]{2017A&A...604A.130K,2019ApJ...870...55G,2019ApJ...883...20G}.
The continuous driving at the loop footpoint changes the 
equilibrium compared with the non-driven case,
leading to the transition region moving with time.
We thus consider an upper region starting from
$9$Mm as the coronal counterpart
in the following analysis.
The volume-averaged Poynting flux and internal energy
are given by 
\begin{eqnarray}
S(t)=-\displaystyle\frac{1}{V}\int^t_0\int_{A'_1}\displaystyle\frac{1}{\mu_0}\left[(\vec{v}\times\vec{B})\times\vec{B}\right]\cdot{\rm d}\vec{A'}_1{\rm d}t,
\label{eq_S_in}
\end{eqnarray}
\begin{eqnarray}
I(t)=\displaystyle\frac{1}{V}\int_{V}\displaystyle\frac{p}{\gamma-1}{\rm d}V-\displaystyle\frac{1}{V}\int^t_0\int_{A'_2}\left(\displaystyle\frac{\rho v^2}{2}+\rho\Phi+\displaystyle\frac{\gamma p}{\gamma-1}\right)\vec{v}\cdot{\rm d}\vec{A'}_2{\rm d}t,
\label{eq_I_in}
\end{eqnarray}
where $V$ represents the volume of the concerned domain
($-6{\rm Mm}\le x\le6{\rm Mm},-3{\rm Mm}\le y\le3{\rm Mm},9{\rm Mm}\le z\le100{\rm Mm}$).
${\rm d}\vec{A}'_1$ (${\rm d}\vec{A}'_2$) is the normal vector 
of the bottom (lateral) surface.
We also incorporate the energy fluxes
of lateral boundaries into the internal energy variations, 
following a similar procedure
in \cite{2019A&A...623A..53K}.
The input energy is defined as the Poynting flux
of the bottom interface.
As aforementioned,
an enthalpy flux can be seen through the bottom boundary 
($z=9$Mm) of the concerned domain.
However,
this enthalpy flux is much smaller than the Poynting flux
here, 
thus has been neglected when considering the input
energy from the bottom.
Figure~\ref{fig_energy} shows the volume averaged
energy density changes relative
to the initial state, namely $S(t)-S(0)$ and $I(t)-I(0)$.
In the non-driven model,
the internal energy ($I_0$) drops with time
due to the thermal conductive loss in the corona,
while the internal energy variation in our driven model ($I_4$) 
stays positive before $120$min and approaches zero 
after about $120$min.
This means that the internal energy first increases and then asymptotically approaches the initial equilibrium state
by the end of the simulation.
Therefore,
the heating induced by waves can overcome or balance the 
thermal conductive loss in the corona,
for the duration of the driving in the current model.

Heating efficiency can be evaluated by considering the 
ratio between the relative internal energy increase and 
the input energy.
Figure~\ref{fig_energy} also shows the difference between 
the driven and non-driven models in the same 
domain.
From the Poynting flux and the difference in internal energy changes,
we can compute the heating efficiency of the driven model.
A rapid increase in Poynting flux before 40min can be 
observed,
leading to an average input energy flux of about 
$78.1{\rm Jm^{-2}s^{-1}}$ over the considered coronal 
counterpart of the loop.
The average internal energy increasing rate is about 
$32.8{\rm Jm^{-2}s^{-1}}$,
leading to a heating rate of $42\%$.
In previous studies
\citep[e.g.,][]{2019ApJ...870...55G, 2019FrASS...6...38K},
we report the expansion of the loop cross-section when 
small scales and turbulence developed,
inducing a decrease in the local magnetic field.
This decrease can also be traced down from the corona
to lower heights with decreasing strength.
The variations of the magnetic field in the lower interface
between the corona and the transition region cause
a slower increase of the input energy from $t\sim50$min,
leading to an average 
input energy flux of about $30.7{\rm Jm^{-2}s^{-1}}$
from $50{\rm min}\le t\le100{\rm min}$.
The corresponding internal energy growth rate is 
$16.8{\rm Jm^{-2}s^{-1}}$ and
the heating rate increases to $55\%$.
In the last 20 minutes of the simulation,
the loop is fully evolved and the expansion is saturated,
we see the input energy flux increase to 
$45.1{\rm Jm^{-2}s^{-1}}$.
The corresponding internal energy increasing rate of this
phase is 
$34.5{\rm Jm^{-2}s^{-1}}$,
leading to a heating rate of $76.5\%$.
This shows a significant heating efficiency of the current model.
Comparing with our previous models
\citep[e.g.,][]{2017A&A...604A.130K, 2019A&A...623A..53K, 2019ApJ...870...55G, 2019ApJ...883...20G, 2021ApJ...908..233S},
the presence of thermal conduction allows enthalpy
flows from the lower atmosphere to the corona.
Such flows change the vertical 
density and temperature distribution in the coronal counterpart of the loop (Figure~\ref{fig_TrhoE}).
The energy dissipation at the loop cross-section becomes more
efficient,
as indicated by the temperature increase in Figure~\ref{fig_isosurface}.
Therefore,
the upward evaporation flows lead to a more efficient heating 
to balance the thermal conductive losses
in the current model.

\section{ DISCUSSION AND SUMMARY}
\label{sec_sum}
 
We model a three-dimensional magnetic cylinder in a realistic solar atmosphere from the chromosphere to the corona
with thermal conduction included.
Frequently observed kink oscillations
are excited in this magnetic flux tube.
Based on previous studies,
heating effects are expected due to the presence of 
the Kelvin-Helmholtz instability induced by kink oscillations.
Temperature and internal energy increases are indeed observed
in the coronal counterpart of our driven model,
with respective to a non-driven case.
With the inclusion of thermal conduction,
chromospheric evaporation can be observed,
leading to an increase in temperature and a decrease in density.
Although the spatially averaged density and temperature changes are gentle ($\sim 4\%$ relative to the non-driven case),
the corresponding enthalpy flow from the lower atmosphere
can still influence the energy dissipation in the corona by 
redistributing the vertical density structuring.
The efficient heating in the coronal counterpart of the loop 
can balance the thermal conductive loss in the current model,
and thus maintain the internal energy in the corona.

 In the current simulation,
we adopt an approximate model derived from 
\cite{2002ApJS..142..269A} 
to describe the vertical distribution 
of the solar atmosphere.
This temperature distribution seems a coarser approximation
than the VAL model \citep{1981ApJS...45..635V} or
its upgraded versions \citep{2008ApJS..175..229A}.
Especially, it lacks sufficient details in the lower atmosphere.
Nevertheless,
it describes the main properties of the parameters 
(e.g., temperature and density)
of the chromosphere and the
transition region.
In addition,
note that the initial analytical profiles of temperature and density 
change after relaxation and a new equilibrium is achieved,
as shown in Figure~\ref{fig_init}b.
The initial expressions of the VAL model of the atmosphere 
would also be changed after the relaxation.

Explicit resistivity and viscosity are not included in the
current ideal MHD computation.
In the solar corona,
the magnetic Reynolds number is of the order of $10^{12}$,
leading to an extremely small resistivity if considering typical 
spatial scales and Alfv\'en speed in the corona.
In practical coronal models,
however,
even the numerical resistivity is several orders larger than the 
realistic value.
Given the numerical resolution limitation,
we should treat such models, especially large 3D models,
as an ideal approximation to the real corona.
Nevertheless,
to examine the influence of explicit resistivity/viscosity
on wave energy dissipation is still necessary.
\citet[][]{2017A&A...602A..74H} found that enhanced
resistivity and viscosity that are larger than numerical values 
can suppress the onset of TWIKH rolls,
thus reducing wave energy dissipation.
The influence of explicit dissipation coefficients on wave heating
has been quantitatively examined by 
\citet{2019A&A...623A..53K}.
Comparing with ideal simulations with only numerical dissipation,
the heating effects are enhanced but the heating locations are the same 
in the resistive/viscous models in \citet{2019A&A...623A..53K}.
This confirms that the numerical dissipation can also play
an effective role in such modeling associated with waves
heating effects.

In previous studies \citep[e.g., ][]{1998ApJ...493..474O},
resonant absorption and 
phase mixing are believed to happen in resonant layers
near a loop boundary.
\citet{2016ApJ...823...31C} believed that
the resonant layer is not sustained due to the temporal evolution
of the density gradient induced by Alfv\'en waves. 
In our current results,
however,
TWIKH rolls spread over the whole cross section of the loop,
associated with chromospheric evaporation not only happening 
near the boundary layer,
but rather in the entire loop.
Nevertheless,
due to the slight density and temperature changes,
the evaporation has no significant 
effect on the resonance conditions of the current model.

In this simulation we are continuously driving waves at the eigenfrequency 
of the standing mode, 
and so the resultant evaporations may be viewed as an upper limit to the evaporations which would 
result from realistic, 
small amplitude (decayless) kink waves which are not so carefully driven. 
The limitation of current-day 3D MHD simulations having a magnetic Reynolds number several orders 
of magnitude off that for the corona further reinforces our results as an upper limit. 

The realistic thermal conductive loss in the quiet solar corona is
about $200{\rm J~m^{-2}~s^{-1}}$ \citep[][]{1977ARA&A..15..363W}.
The input energy in our current model does not seem to be 
able to complement such a high conductive loss in the real corona.
However,
it should be noted that the magnetic field strength of 
the current model is still vertically uniform.
In reality,
if the loop footpoint is anchored in the photosphere,
the magnetic field should be much (or even several orders of magnitude)
larger than that in the corona.
This probably indicates that the real energy input
from the photosphere is sufficient.
In fact,
observations have confirmed that the energy flux 
is indeed sufficient to balance the energy losses even in the 
active region
\citep[e.g.,][]{2009ApJ...702.1443F}.

We have simulated a more realistic coronal loop by including a dense and cold chromosphere, 
and despite the additional dynamics and energetics, 
it is encouraging to see that the results about wave heating inferred from coronal-only simulations
still seems to hold. 
Examples include the formation of extended TWIKH rolls along the whole loop; 
and the increased efficiency of wave heating once the kinetic energy saturates at the later stages 
of the simulation and a turbulent density profile has developed.

\clearpage
\begin{acknowledgments}
We thank the referee for helpful comments that improved the manuscript. The authors acknowledge the funding from the European Research Council (ERC) under the European Unions Horizon 2020 research and innovation program (grant agreement No. 724326).
  TVD was also supported by the C1 grant TRACEspace of Internal Funds KU Leuven, and a Senior Research Project (G088021N) of the FWO Vlaanderen.
  K.K. acknowledges support by an FWO (Fonds voor Wetenschappelijk Onderzoek-Vlaanderen) postdoctoral fellowship (1273221N).
  Y.G. acknowledges the support from the China Scholarship Council (CSC) under file No. 202206010018.
\end{acknowledgments}

\clearpage
\bibliographystyle{apj}
\bibliography{kink_heating}


\end{document}